\documentclass[11pt,a4paper]{amsart}

\pdfoutput=1

\usepackage{geometry}                % See geometry.pdf to learn the layout options. There are lots.
\usepackage{graphicx}
\usepackage{amssymb}
\usepackage{epstopdf}
\usepackage{color}
\usepackage{float}
\usepackage{caption}
\usepackage[table]{xcolor}

\usepackage{url}

\RequirePackage[OT1]{fontenc}
\RequirePackage{amsthm,amsmath}
\usepackage{amsmath, amssymb, amsfonts} % for math symbols like \Sigma
\usepackage{booktabs}  % for tables
\usepackage{graphicx} % for figures

\theoremstyle{plain}% Theorem-like structures provided by amsthm.sty
\newtheorem{theorem}{Theorem}[section]

\theoremstyle{definition}

\theoremstyle{remark}

\let\hat\widehat
\let\tilde\widetilde

\renewcommand\P{\mathsf{P}}

\DeclareMathOperator*{\di}{\mathrm{d}\!}

\begin{document}

\title[Measuring the temporal stability of human mobility patterns]{A statistical framework for measuring the temporal stability of human mobility patterns}\thanks{CONTACT A. Dobra. Email: adobra@uw.edu}

\author{Zhihang Dong\textsuperscript{*,$\dagger$}, Yen-Chi Chen\textsuperscript{*} and Adrian Dobra\textsuperscript{*}}
\address{\textsuperscript{*} Department of Statistics, University of Washington, Seattle, WA, USA; \textsuperscript{$\dagger$} Department of Sociology, University of Washington, Seattle, WA, USA}

\maketitle

\date{\today} 

\begin{abstract}
Despite the growing popularity of human mobility studies that collect GPS location data, the problem of determining the minimum required length of GPS monitoring has not been addressed in the current statistical literature. In this paper we tackle this problem by laying out a theoretical framework for assessing the temporal stability of human mobility based on GPS location data. We define several measures of the temporal dynamics of human spatiotemporal trajectories based on the average velocity process, and on activity distributions in a spatial observation window. We demonstrate the use of our methods with data that comprise the  GPS locations of 185 individuals over the course of 18 months. Our empirical results suggest that GPS monitoring should be performed over periods of time that are significantly longer than what has been previously suggested. Furthermore, we argue that GPS study designs should take into account demographic groups.
\end{abstract}

\tableofcontents

\section{Introduction}

Recent developments on global positioning systems (GPS) for wearable technology such as smartphones have drawn a great amount of interest from scientists studying the effects of environmental influences on different population groups \cite{RN143,RN144,RN146,RN149,RN145,zenk2011activity,RN147,wiehe2013adolescent,entwisle2007putting}. A recent article \cite{mazimpaka2016trajectory}  documents more than 100 studies from 20 disciplines that collect and analyze human time-stamped GPS location data. This type of data is key for learning about the places where people routinely spend their time during activities of daily living in order to establish their relationship with socio-economic outcomes, crime victimization, and physical and mental well-being. There have been extensive studies on the social stratification of mobility, such as health disparities of different neighborhoods, mental health, and substance abuse intervention \cite{elgethun2007comparison,vazquez2013using,wiehe2013adolescent}, on the assessment of human spatial behavior and spatiotemporal contextual exposures \cite{RN144,RN146,RN149}, on the characterization of the relationship between geographic and contextual attributes of the environment (e.g., the built environment) and human energy balance (e.g., diet, weight, physical activity) \cite{RN145,zenk2011activity},  on the study of segregation, environmental exposure, and accessibility in social science research \cite{RN147}, or on the understanding of the relationship between health-risk behavior in adolescents (e.g., substance abuse) and community disorder \cite{wiehe2013adolescent,basta-et-2010,RN153}.

Notwithstanding a general consensus across disciplines about the tremendous potential of GPS location data for studying human mobility, very little is currently known about how long a GPS study should last. There is an inherent trade-off between collecting location data from people for longer vs. shorter periods of time. Recording more GPS locations yields more information about the locations where an individual spends their time, as well as about the frequency, duration and timing of their visits to these places. However, an individual's participation in a GPS study comes with burdens that often become significant if accumulated over longer periods of time: the individual needs to carry the device recording the data (a GPS tracker) everywhere they go, and needs to make sure the device is properly charged at all times and functions properly. Until recently, most GPS study designs stipulated mandatory regular visits to project coordination sites to download data from the location trackers, to replace batteries, and replace the GPS tracking devices that were lost or were malfunctioning. While some of these issues have been addressed by using specialized apps on smartphones to collect GPS data and wirelessly transmit them into secure cloud databases, the costs of distributing smartphones to study participants, data plans, software development, and cloud computing are quite significant. In addition, there are important privacy considerations related to recording locations that might sensitive for study participants for long periods of time. For these reasons, it is desirable to design GPS studies that are as short as possible to reduce the costs of the projects and the burden of study participants, while in the same time still providing guarantees that sufficient location data have been collected to properly address the research aims.

Despite the constant growth in the number of human mobility studies that collect GPS location data in the last 20 years, the question about the determination of the amount of time of GPS monitoring has not been asked until recently  \cite{zenk2018many}. In this paper, the authors argue that an effective GPS study should last until a minimum of 14 to 15 days of valid GPS data have been collected. While this finding is relevant for numerous research groups that, in the past, have designed GPS studies with a duration of 7 days (see \cite{zenk2018many} and the references therein), two weeks seems to severely underestimate the duration of other, more recent, GPS studies whose duration is significantly longer. For example, \cite{byrnes-et-2017} and \cite{morrison-et-2019} represent studies that tracked adolescents in the San Francisco Bay area for one month. Another study \cite{duncan-et-2019} employs a more complex three site design that comprises five assessments that take place every six months over two years of follow-up for participants enrolled in Chicago, and three assessments that take place every six months over one year of follow-up for participants enrolled in Jackson and New Orleans. During each assessment, participants wear a GPS tracker for two weeks. Thus this study \cite{duncan-et-2019} records GPS locations for a total of 10 weeks and 6 weeks, respectively, but splits the period of observation into several contiguous two week periods of GPS monitoring. These longer periods of observation time were suggested in \cite{lee2016activity} who found 17 weeks to be an adequate period of time to monitor human mobility based on geotagged social media data.

In this paper we lay out a theoretical framework for assessing the temporal stability of human mobility based on GPS location data. Such a framework is missing from the current statistical literature. Previous work \cite{zenk2018many,lee2016activity} on the assessment of the duration of GPS observation periods is based on empirical findings, and lack any theoretical underpinnings. We address this gap by introducing several measures of the temporal dynamics of spatiotemporal trajectories of individuals. We illustrate the use of these measures with publicly available data from a study that recorded GPS locations of 185 individuals that live in a city in Switzerland over the course of 18 months.

\section{Methods}	\label{sec::method}

The spatiotemporal trajectory of an individual in a reference time frame $[t_{\min},t_{\max}]$ and spatial observation window $\mathcal{W}\subset \mathbb{R}_{+}^2$ is a curve
\begin{eqnarray} \label{eq:trajectory}
 X^{[t_{\min},t_{\max}]} & = & \{ X(t) = (x_1(t), x_2(t)): t\in [t_{\min},t_{\max}]\} \subseteq \mathcal{W},
\end{eqnarray}
\noindent where $x_1(\cdot)$ and $x_2(\cdot)$ represent the longitude and latitude coordinates, respectively, and $X(t)$ is the location visited by this individual at time $t$. We assume that this curve is smooth: $x_1(\cdot)$ and $x_2(\cdot)$ have continuous derivatives. The length of the curve in Eq. \eqref{eq:trajectory} is defined as \cite{courant-john-1991}:
\begin{eqnarray} \label{eq:curvelength}
  {\sf L}(X^{[t_{\min},t_{\max}]}) & = & \int\limits_{t_{\min}}^{t_{\max}} \sqrt{\left( \frac{\di x_1(t)}{\di t} \right)^2 + \left( \frac{\di x_2(t)}{\di t} \right)^2} \di t.
\end{eqnarray}
The complete trajectory $X^{[t_{\min},t_{\max}]}$ is never observed in the real world. Instead, $n$ observation times $t_1,\ldots,t_n$ are sampled from a distribution on $[t_{\min},t_{\max}]$ with density $\rho(\cdot)$, and the corresponding locations $X(t_1),\ldots,X(t_n)$ on the curve $X^{[t_{\min},t_{\max}]}$ are recorded. These locations are realizations of a random variable $X(T)$ where $T\sim q(\cdot)$. Ideally we would like $T$ to follow a uniform distribution to have the same chance of recording a visited location anywhere in the reference time frame $[t_{\min},t_{\max}]$. Due to technological limitations (e.g., GPS devices running out of power), heterogeneous built environments that prevent GPS devices to obtain a location (e.g., skyscapers in downtown areas or buildings without windows and WIFI coverage), or human behavioral factors (e.g., individuals turning off their GPS devices around certain locations sensitive to them) the distribution of $T$ can be far from the uniform distribution.

We assume that GPS positional data from $K$ study participants were recorded. We denote by $X^{[t_{\min},t_{\max}]}_k = \{ X_k(t): t\in [t_{\min},t_{\max}]\}$ the unobserved spatiotemporal trajectory of the $k$-th study participant. The observation times in the reference time frame $[t_{\min},t_{\max}]$ can vary between study participants. The GPS data for the $k$-th study participant are the time stamped longitude and latitude locations:
\begin{eqnarray}\label{eq:positionaldata}
  \{ X_{k,i} = X_k(t_{k,i}): i=1,\ldots,n_k\},
\end{eqnarray}
\noindent where $n_k\ge 1$, the time $t_{k,i}$ was sampled from a distribution with density $\rho_k(\cdot)$ independently of the rest of the observation times, and $t_{\min}\le t_{k,1}\le \ldots t_{k,n_k}\le t_{\max}$. Here $t_{k,i}$ represents the time when the $i$-th location of study participant $k$ was recorded. Our framework allows for the possibility of having different reference time frames for various groups of study participants.

\subsection{Measuring the temporal stability of human mobility patterns}

One possible measure of the dynamics of the spatiotemporal trajectory $X^{[t_{\min},t_{\max}]}$ is the average velocity $V(\tau)$ at time $\tau$ which is a function $V(\tau)$ of the length of the subcurve $X^{[t_{\min},t_{\min}+\tau]}$ of $X^{[t_{\min},t_{\max}]}$ from Eq. \eqref{eq:trajectory}:
\begin{eqnarray} \label{eq:velocity}
 V(\tau) & = & \frac{1}{\tau} {\sf L}(X^{[t_{\min},t_{\min}+\tau]}), 
\end{eqnarray}
\noindent for $\tau \in (0,t_{\max}-t_{\min}]$ and $V(0) = 0$. A sample estimator of the average velocity for the $k$-th study participant is
\begin{eqnarray} \label{eq:velocitysample}
 \widehat{V}_k(\tau) & = & \frac{1}{\tau}\sum_{\{i: t_{k,i+1}\le \tau\}} \|X_{k,i+1}- X_{k,i}\|.
\end{eqnarray}
where $\|X_{k,i+1}- X_{k,i}\|$ represents an estimate of the distance traveled between times $t_{k,i}$ and $t_{k,i+1}$. In what follows we will assume that study participants traveled in a straight line or ``as the crow flies" between two consecutive observed GPS locations. This is the simplest assumption one can make which leads to an easy way of calculating Great Circle (WGS84 ellipsoid) distances between two spatial locations \cite{bivand-et-2013}. However, this assumption underestimates actual distances traveled, and consequently underestimates the average velocity. More accurate approximations of distances traveled can be defined based on the shortest distances between two locations on a road network that spans the spatial observation window $\mathcal{W}$. Calculating distances based on a road network is more complex than calculating straight line distances, and involves significant GIS work since the maximum speed of travel on different segments of road needs to be taken into account \cite{dobra-et-2015}. Nevertheless, as the span of time between two consecutive observed locations becomes shorter, the difference between the road network and straight line distances decrease.

More generally, consider a stochastic process $Z = \{ Z(\tau):\tau  \in [0,t_{\max}-t_{\min}]\}$, where $Z(\tau)$ is a mapping $f(\cdot)$ of the subcurve $X^{[t_{\min},t_{\min}+\tau]}$ into $\mathbb{R}_{+}$. The mapping $f(\cdot)$ is chosen such that $\lim\limits_{\tau \rightarrow (t_{\max}-t_{\min})} Z(\tau) = Z(t_{\max}-t_{\min})$. We define the absolute percentage error (APE, henceforth) $\phi(Z;\tau)$ which measures the error made when approximating $Z(t_{\max}-t_{\min})$ with $Z(\tau)$ for $\tau \in [0,t_{\max}-t_{\min}]$:
\begin{eqnarray*}
 \phi(Z;\tau) & = & \frac{|Z(\tau) - Z(t_{\max}-t_{\min})|}{Z(t_{\max}-t_{\min})}.
\end{eqnarray*}
We quantify the temporal stability of the process $Z$ by introducing a related process called the last crossing time process ${\sf LCT}_{Z} = \{ {\sf LCT}_{Z}(\gamma) : \gamma\ge 0\}$, where
\begin{eqnarray} \label{eq:lct}
 {\sf LCT}_{Z}(\gamma)  = \max \left\{\tau \in [0,t_{\max}-t_{\min}]: \phi(Z;\tau) > \gamma\right\}.
\end{eqnarray}
\noindent In Eq. \eqref{eq:lct}, ${\sf LCT}_{Z}(\gamma)$ is the last time when the APE made when $Z(t_{\max}-t_{\min})$ is approximated with $Z(\tau)$ is above a threshold $\gamma$. The last crossing time is well defined since $\lim\limits_{\tau \rightarrow (t_{\max}-t_{\min})} \phi(Z;\tau) = 0$.

\begin{figure}[h]
  \centering
  \includegraphics[height=2.5in]{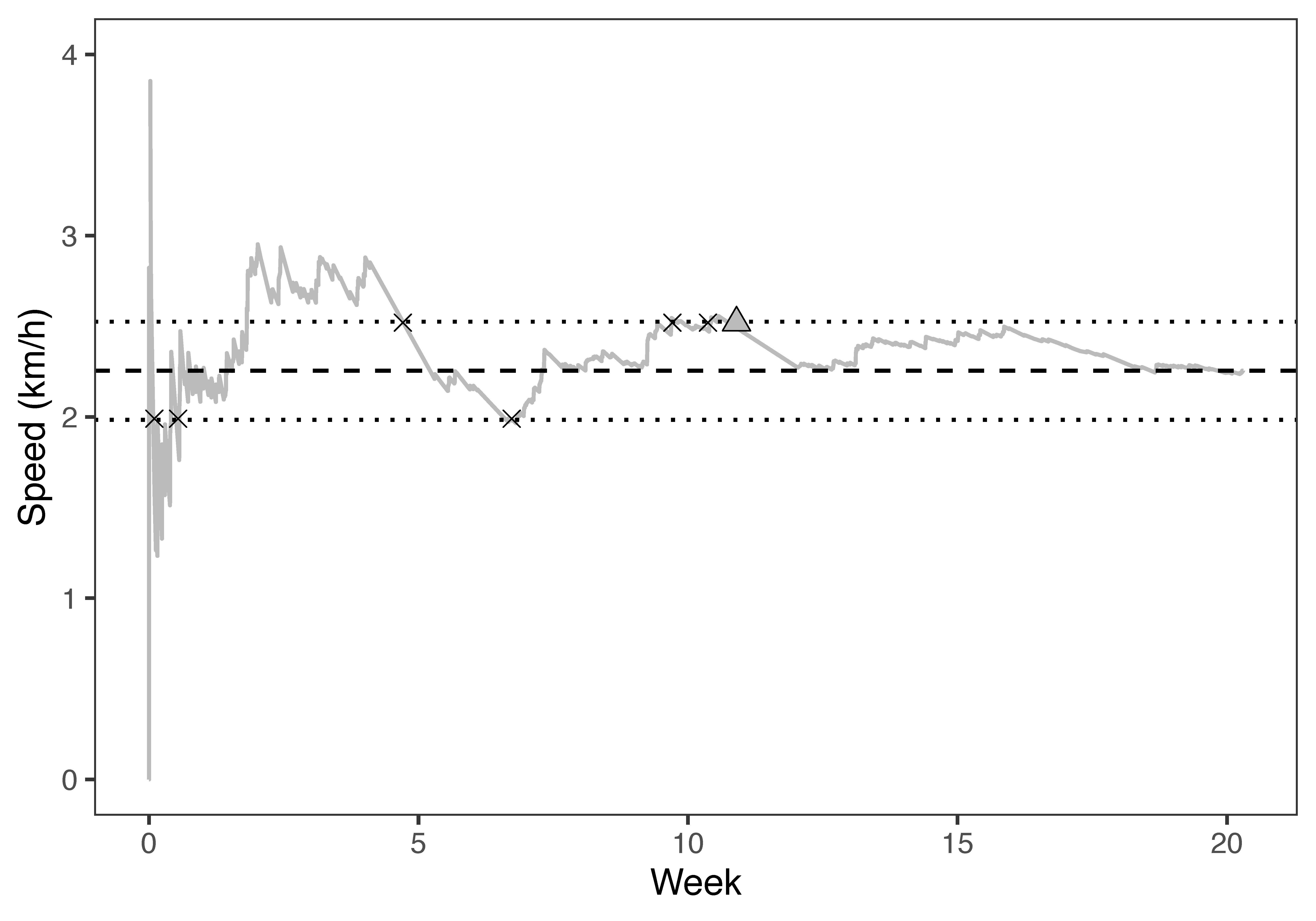}
  \caption{Estimate of the average velocity (gray curve) of an individual in the MDC data over $t_{\max} = 21$ weeks. The dashed line indicates the value of $\widehat{V}(t_{\max})$, and the two dotted lines represent the lower bound $(1-\gamma)\widehat{V}(t_{\max})$ and the upper bound $(1+\gamma)\widehat{V}(t_{\max})$ for $\gamma=0.1$. These bounds correspond with times $\tau$ for which the APE $\phi(V;\tau)\le \gamma$. The crosses denote the times $\tau$ for which  $\phi(V;\tau) =  \gamma$. The last crossing time for $\gamma=0.1$ is marked with a triangle, and occurs at the end of week 10.
}
\label{fig2}
\end{figure}

Consider the process $Z_k= \{ Z_k(\tau):\tau  \in [0,t_{\max}-t_{\min}]\}$ associated with the $k$-th study participant, $Z_k(\tau)=f\left(X_k^{[t_{\min},t_{\min}+\tau]}\right)$, and let $\widehat{Z}_k$ be its sample estimator based on the positional data in Eq. \eqref{eq:positionaldata}. The average velocity in Eq. \eqref{eq:velocity} and its sample estimator in Eq. \eqref{eq:velocitysample} are examples of processes $Z_k$ and $\widehat{Z}_k$. A sample estimator of the last crossing time ${\sf LCT}_{Z_k}(\gamma)$ is
\begin{equation}\label{eq::LCT}
\widehat {\sf LCT}_{Z_k}(\gamma) = \max_{i=1,\ldots,n_k} \left\{t_{k,i}-t_{\min}: \phi(\widehat{Z}_k; t_{k,i}-t_{\min}) > \gamma\right\}.
\end{equation}
We note that $\widehat{Z}_k(\tau)$ in the APE $\phi(\widehat{Z}_k; \tau)$ is determined based on the locations recorded for the $k$-th study participant before time $\tau$: $\{X_{k,i}: t_{\min}\le t_{k,i}\le \tau\}$. As an illustration, Figure~\ref{fig2} shows estimates of the average velocity of an individual in the MDC data, together with the last crossing time estimate at $\gamma = 0.1$.

The last crossing time of the APE associated with a process that is a function of the spatiotemporal trajectory of a study participant represents a measure of this individual's mobility. Study participants that have more irregular mobility patterns (e.g., regular travel to locations at various distances from the individual's residence that change after a few days or weeks) are expected to have larger last crossing times compared to study participants that travel to the same locations each week. An example individual with a very regular mobility pattern that travels every day from his home to his office and back by following the same route, and goes nowhere else will record an APE equal to $0$ after one day which leads to last crossing times of less than one day in Eq. \eqref{eq::LCT}.

Previous work \cite{zenk2018many} on the temporal stability of spatiotemporal trajectories has used the mean absolute percentage error (MAPE) which is the average of the APE across study participants:
\begin{eqnarray}
 \overline{\phi}_{K}(\tau)  = \frac{1}{K} \sum\limits_{k=1}^K \phi(\widehat{Z}_k;\tau).
\end{eqnarray}

We define two measures of the overall temporal stability of the spatiotemporal trajectories of multiple study participants. The first overall measure is the last crossing time process ${\sf LCT}_{\overline{\phi}_K} = \{ {\sf LCT}_{\overline{\phi}_K}(\gamma) : \gamma\ge 0\}$ of the MAPE process $\overline{\phi}_K = \{\overline{\phi}_K(\tau):\tau  \in [0,t_{\max}-t_{\min}]\}$. We refer to this measure as ${\sf LCT-MAPE}(Z)$. The second overall measure is defined as the average of the last crossing times of the APE of $\widehat{Z}_k$ for $k=1,\ldots,K$, i.e. $\overline{\sf LCT}_K = \{ \overline{\sf LCT}_K(\gamma) : \gamma\ge 0\}$ where
\begin{eqnarray*}
 \overline{\sf LCT}_K(\gamma) & = & \frac{1}{K}  \sum\limits_{k=1}^K {\sf LCT}_{Z_k}(\gamma).
\end{eqnarray*}
\noindent We denote this second measure by ${\sf \overline{LCT}-APE}(Z)$. These two measures are the same only if they are calculated for a single study participant ($K=1$). They are useful for comparing the temporal regularity of mobility patterns of groups of study participants (e.g., younger vs. older individuals, men vs. women, high SES vs. low SES).

\subsection{The activity distribution of human mobility patterns}

The average velocity associated with the spatiotemporal trajectory of an individual does not provide any information about the spatial configuration of locations visited. Consider two example individuals that drive without stopping with the same speed for a long period of time. The first example individual drives back and forth between two places $A_1$ and $A_2$. The second example individual drives in a cycle from a place $A_1$ to another place $A_2$, then to places $A_3$ and $A_4$, then back to place $A_1$. Since the spatiotemporal trajectory of the second individual involves two additional places, more sample locations will be needed to understand the mobility pattern of the second individual compared to the mobility pattern of the first individual. However, the mobility patterns of these two example individuals will be indistinguishable based on the last crossing time process associated with their average velocity processes. We address this issue by introducing a distribution of the locations visited by an individual. 

We assume that the observation window $\mathcal{W}$ is partitioned into a set of grid cells $\mathcal{G} = \{ G_1,\ldots,G_N\}$. Each location $X(t)$ on the curve $X^{[t_{\min},t_{\max}]}$ representing the spatiotemporal trajectory of an individual is mapped into a grid cell $G(t)\in \mathcal{G}$. The observed locations for this individual mapped into $\mathcal{G}$ are the sequence of grid cells $g_1=G(t_1),\ldots,g_n=G(t_n)$ that are realizations of a random variable $G(T)$ where $T$ is a random variable on $[t_{\min},t_{\max}]$ with a distribution with density $\rho(\cdot)$. 

We define the activity distribution $\pi = (\pi_1,\ldots,\pi_N)$ over the grid cells $\mathcal{G}$. Here $\pi_j$ represents the proportion of time in $[t_{\min},t_{\max}]$ spent by an individual in cell $G_j\in \mathcal{G}$. We assume that $T$ follows a uniform distribution on $[t_{\min},t_{\max}]$, and define:
\begin{equation} \label{eq:activityunif}
 \pi_j = \P(G(T)=G_j), \quad \mbox{for } j=1,\ldots,N.
\end{equation}
The activity distributions associated with the two example individuals we introduced earlier can differentiate between their mobility patterns if the grid cells in which $A_3$ and $A_4$ do not coincide with the grid cells of $A_1$ and $A_2$, and will show that the first example individual did not spend any time in the grid cells associated with $A_3$ and $A_4$. To employ activity distributions we need to have a method for recovering them from the available data.

The simplest estimator $\hat{\pi} = (\hat{\pi}_{1},\ldots,\hat{\pi}_{N})$ of the activity distribution $\pi$ is based on the relative frequency of visitation of the grid cells $\mathcal{G}$: 
\begin{equation*}
   \hat{\pi}_j = \frac{1}{n}\sum^n_{i=1}\mathbf{1}(g_i = G_j), \quad \mbox{for } j=1,\ldots,N.
\end{equation*}
However, this estimator of $\pi$ is reasonable only if $T$ follows a uniform distribution as in Eq. \eqref{eq:activityunif}. When $T$ follows an arbitrary distribution with density $\rho(\cdot)$, a better approach is to use a weighted average estimator $\tilde{\pi} = (\tilde{\pi}_1,\ldots,\tilde{\pi}_N)$ where:
\begin{equation}
   \tilde{\pi}_j = \frac{\sum^n_{i=1}\rho^{-1}(t_i) \mathbf{1}(g_i=G_j)}{\sum^n_{\ell=1}\rho^{-1}(t_\ell)}, \quad \mbox{for } j=1,\ldots,N.
   \label{eq::wa}
\end{equation}
Although this estimator can be shown to be statistically consistent, it requires knowledge of the density $\rho(\cdot)$. There are many methods for estimating $\rho(\cdot)$ from the data such as histograms or kernel density estimators \cite{wasserman-nonparametric}. We suggest using an estimation method that assumes that the distribution of $T$ is approximated by a piecewise uniform distribution. We take $t_0=t_{\min}$ and $t_{n+1} = t_{\max}$. If $T$ is approximately uniform in $[t_{i-1},t_{i+1}]$ for $i=1,\ldots,n$, then $\rho^{-1}(t_i) \approx t_{i+1}-t_{i-1}$. This is a reasonable assumption if the times when locations are collected are roughly equally spaced in time (e.g., a location is collected every 10 minutes) since the mean of $t_i$ is $(t_{i+1}-t_{i-1})/2$. Thus an estimator of $\rho(\cdot)$ is
\begin{equation*}
%    \label{eq9}
    \hat{\rho}(t_i) = \frac{\omega({t_i})}{\sum^n_{\ell=1}\omega(t_\ell)}, \textrm{ } \omega(t_i) = \frac{1}{t_{i+1}-t_{i-1}}, \quad \mbox{for } i=1,\ldots,n.
\end{equation*}
The weighted average estimator from Eq. \eqref{eq::wa} becomes
\begin{equation}
	\label{eq10}
    \begin{split}
        \hat{\pi}_{o,j} &= \frac{\sum^n_{i=1}{\omega}^{-1}(t_i)\mathbf{1}(g_i=G_j)}{\sum^n_{\ell=1}\omega^{-1}(t_\ell)}\\
        & = \frac{\sum^n_{i=1}(t_{i+1}-t_{i-1})\mathbf{1}(g_i = G_j)}{t_{\max}-t_{min}+t_n-t_1}, \quad \mbox{for } j=1,\dots,N.
    \end{split}
\end{equation}
We call $\hat{\pi}_o = (\hat{\pi}_{o,1},\ldots,\hat{\pi}_{o,N})$ the ordinary proportional time estimator of the activity distribution $\pi$. This estimator relies on the assumption that the length of the time intervals in which an individual transitions between two grid cells is added to the time spent in both the grid cell they leave from, and the grid cell they arrive in. More specifically, assume that the consecutive observation times $t_i$ and $t_{i+1}$ are such that $g_i\ne g_{i+1}$. Then $\hat{\pi}_o$ allocates $(t_{i+1}-t_i)$ to the total time spent in both $g_i$ and $g_{i+1}$. 

We introduce a second estimator $\hat{\pi}_c = (\hat{\pi}_{c,1},\ldots,\hat{\pi}_{c,N})$ of the activity distribution $\pi$:
\begin{equation}
\label{eq11}
\hat{\pi}_{c,j} = \frac{\sum^n_{i=2}(t_i-t_{i-1})\mathbf{1} (g_i = g_{i-1} = G_j)}{\sum^n_{i=2}(t_i-t_{i-1}) \mathbf{1} (g_i = g_{i-1})},\quad \mbox{for } j=1,\ldots,N.
\end{equation}
We call $\hat{\pi}_c$ the conservative proportional time estimator. This estimator is more conservative than the ordinary proportional time estimator $\hat{\pi}_o$ from Eq. \eqref{eq10} in the sense that any time interval defined by consecutive observation times $t_i$ and $t_{i+1}$ such that $g_i\ne g_{i+1}$ is ignored. That is, the time spent in a grid cell is calculated only based on time intervals in which an individual is known to have remained in that cell.

We show two important properties of the ordinary and the conservative proportional time estimators. First, we prove that both estimators are asymptotically equivalent. Second, we prove that both estimators are statistically consistent, that is, they will eventually recover the true activity distribution $\pi$ if sufficient location data are available. These properties rely on the assumptions (S1), (S2) and (S3) below:

\begin{itemize}
\item[(S1)] The length of the time intervals between consecutive observation times $\max\limits_{i=1,\ldots,n-1} |t_{i+1} - t_i| \to 0$ as the sampling rate $n\rightarrow\infty$.
\item[(S2)] The sampling period is such that $t_1\rightarrow t_{\min}$ and $t_n\rightarrow t_{\max}$ when $n\rightarrow\infty$. 
\item[(S3)] The number of transitions between grid cells is finite, i.e., there exists $M<\infty$ such that 
$\sum_{t\in [t_{\min},t_{\max}]}\mathbf{1}(G(t_+)\neq G(t_-)) \leq M$,
where $G(t_-)$ and $G(t_+)$ are the left and right limits of $G(\cdot)$ at $t$.
\end{itemize}

Assumptions (S1) and (S2)  describe the meaning of asymptotics in our context. They imply that 
the observation times $t_1,\ldots, t_n$ will eventually be dense in the reference time  frame, i.e., there will not exist a fixed region of $[t_{\min},t_{\max}]$ without any observation times when $n\rightarrow \infty$. Assumption (S3) requires that the spatiotemporal trajectory $X^{[t_{\min},t_{\max}]}$ is sufficiently smooth such that it will not jump between grid cells infinitely often. 

\begin{theorem}[Asymptotic Equivalence Rule with Large Sampling Rate]
Under assumptions (S1), (S2) and (S3), the ordinary proportional time estimator $\hat{\pi}_o$ from Eq. \eqref{eq10} and the conservative proportional time estimator $\hat{\pi}_c$ from Eq. \eqref{eq11} are asymptotically the same.
\label{thm::equiv}
\end{theorem}

The proof of this result is given in Appendix \ref{proof::equiv}. We can also show that  the same assumptions imply that the two estimators are statistically consistent. 

\begin{theorem}[Convergence Rule with Large Sampling Rate]
Under assumptions (S1), (S2) and (S3), the ordinary proportional time estimator $\hat{\pi}_o$ from Eq. \eqref{eq10} and the conservative proportional time estimator $\hat{\pi}_c$ from Eq. \eqref{eq11} converge to the true activity distribution $\pi$ from Eq. \eqref{eq:activityunif}.
	\label{thm::convergencerule}
\end{theorem}

The proof of this result is given in Appendix \ref{proof:convergencerule}.

\subsection{Measuring the temporal stability of human activity distributions}

We are interested in determining the temporal stability of the activity distribution of an individual. We assume that the reference time frame $[t_{\min},t_{\max}]$ is divided into $D_{\max}$ time periods of equal lengths (e.g., days or weeks). We denote by $\pi^{(d)}$ the activity distribution from Eq. \eqref{eq11} associated with time period $D$, $D=1,\ldots,D_{\max}$. Then $\pi^{(D)}$ can be viewed as an $N$-dimensional random vector whose distribution reflects the variability from time period to time period of the individual's mobility patterns. With this understanding, we are interested in determining the expectation $\bar \pi = \mathbb{E}(\pi^{(D)})$. We call $\bar \pi$ the time period activity distribution (e.g., daily or weekly activity distribution). The $j$-th component of $\bar \pi$ is interpreted as the average proportion of time spent by the individual in grid cell $G_j$ in a given time period (a day or a week).

A simple estimator of $\bar \pi$ is 
\begin{equation} \label{eq:piD}
 \hat {\bar {\pi}}(D) = \frac{1}{D}\sum_{d=1}^{D}\hat \pi^{(d)},\quad \mbox{for } D=1,\ldots,D_{\max},
\end{equation}
where $\hat \pi^{(d)}$ is the ordinary proportional time estimator $\hat{\pi}_o$ from Eq. \eqref{eq10} or the conservative proportional time estimator $\hat{\pi}_c$ from Eq. \eqref{eq11}.

Because $\hat {\bar {\pi}}({D})$ is a consistent estimator of $\bar \pi$, the error we make when approximating $\bar \pi$ with $\hat {\bar {\pi}}({D})$ decreases as we observe the spatiotemporal trajectory of the individual for a larger number of time periods $D_{\max}$. We define the last crossing time of the sequence of estimators $\{ \hat {\bar {\pi}}(D): D=1,\ldots,D_{\max}\}$ as follows:
\begin{equation}
\begin{aligned}
\hat {\sf LCT}_{\sf dist}(\gamma) &= \max\limits_{D=1,\ldots,D_{\max}} \left\{D: \|\hat {\bar {\pi}}({D}) - \hat {\bar {\pi}}({D_{\max}})\|_1 > \gamma\right\},
\end{aligned}
\label{eq::LCT::dis}
\end{equation}
where $\|v\|_1$ is the usual $L_1$ norm for a vector $v$, i.e., $\|v\|_1 = \sum_{i} |v_i|$. Note in Eq. \eqref{eq::LCT::dis} we used the fact that $\|\hat {\bar {\pi}}({D})\|_1 = 1$ for any $D$.

The last crossing time in Eq. \eqref{eq::LCT::dis} is a measure of the temporal stability of the entire time period activity distribution $\bar \pi$. Individuals that spend approximately the same amount of time in the same places in every time period need to be observed for a smaller number of time periods to calculate estimator $\hat {\bar {\pi}}(D)$ with the same APE compared to individuals with heterogeneous mobility patterns that spend different amounts of times at locations that change substantially across time periods. Therefore $\hat {\sf LCT}_{\sf dist}(\gamma)$ will be smaller for individuals whose time period to time period mobility changes less, and larger for individuals with irregular mobility patterns.

The disadvantage of using the last crossing time in Eq. \eqref{eq::LCT::dis} as a measure of temporal stability comes from the fact that it gives the same weight to the error made when estimating the proportion of time spent in grid cells in which an individual spends a lot of their time, and to the grid cells in which the individual rarely visits. The number of grid cells with a large proportion of time spent in them is likely significantly smaller than the total number of grid cells $N$ because most people tend to spend time at their residence, to their work place and perhaps in a few other select locations. For this reason, the error made when estimating the proportion of time spent in grid cells with sparse presence could dominate the overall APE of $\hat {\bar {\pi}}(D)$, and lead to larger values of $\hat {\sf LCT}_{\sf dist}(\gamma)$. To remedy this issue, we define a new measure of temporal stability that focuses on the grid cells in which an individual spends larger proportions of time.

We define the ranking time period activity distribution ${\bar r} = ({\bar r}_1,\cdots,{\bar r}_N)$ associated with $\bar{\pi}$ by replacing each component of $\bar{\pi}$ with the sum of those components of $\bar{\pi}$ that are no larger than that component, as follows \cite{chen-2019}:
\begin{equation} \label{eq:rankingdistrib}
 \bar{r}_j = \sum\limits_{l=1}^N \bar{\pi}_l \mathbf{1}( \bar{\pi}_l \le \bar{\pi}_j), \quad\mbox{for } j=1,\ldots,N.
\end{equation}
The $\alpha$-level set ($\alpha\in [0,1]$) of ${\bar r}$ is defined to consist of all the grid cells whose corresponding components in ${\bar r}$ exceed $\alpha$:
\begin{equation} \label{eq:levelset}
 L_{\alpha} = \{ G_j: \bar{r}_j  \ge \alpha\}.
\end{equation}  
It turns out that the  $\alpha$-level set covers grid cells whose total sum of components of $\bar{\pi}$ is larger than $1-\alpha$:
\begin{equation*}
 \sum_{G_j\in L_\alpha} {\bar \pi}_j \geq 1-\alpha.
\end{equation*}
Levels sets have an easy to understand interpretation: for a given level $\alpha$, say $\alpha = 0.7$, all the grid cells with a ranking time period activity distribution above $0.7$ will jointly cover at least $(1-0.7)\cdot 100 = 30$\% of the time in the time period. Values of $\alpha$ closer to $1$ lead to level sets $L_{\alpha}$ with a smaller coverage that comprise only the grid cells in which the individual spends the largest amounts of time. Values of $\alpha$ close to $0$ lead to level sets $L_{\alpha}$ with a larger coverage that comprise the majority of grid cells the individual spent time in.

Let $\hat {\bar r}(D)$ be the ranking distribution of the estimator $\hat {\bar {\pi}}(D)$ of $\bar{\pi}$ in Eq. \eqref{eq:piD}, and $L_{\alpha}(D)$ be the $\alpha$-level set associated with $\hat {\bar r}(D)$ as in Eq. \eqref{eq:levelset}. Given a level $\alpha\in [0,1]$ and a stability threshold $\gamma>0$, we define the last crossing time of the sequence of level sets $\{ L_\alpha(D): D=1,\ldots,D_{\max}\}$ as follows:
\begin{equation}
\hat {\sf LCT}_{\sf level, \alpha}(\gamma) = \max\limits_{D=1,\ldots,D_{\max}}  \left\{D: \frac{\|L_\alpha(D) \triangle L_\alpha(D_{\max})\|}{\|L_\alpha(D_{\max})\|} > \gamma\right\},
\label{eq::LCT::LV}
\end{equation}
where $\triangle$ denotes the symmetric difference of two sets, and $\|\cdot\|$ denotes the number of elements in a set.

The LCT of the level sets from Eq. \eqref{eq::LCT::LV} is a measure of temporal stability of the time period activity distribution $\bar{\pi}$ that takes into account only the error made when estimating the time spent in the grid cells in which an individual spent most of their time. For the same value of $\gamma$, $\hat {\sf LCT}_{\sf level, \alpha}(\gamma)$ is decreasing as the level $\alpha$ is increasing. 

\section{Application}

The data we analyze comes Nokia's Mobile Data Challenge (MDC) \cite{kiukkonen-et-2010,laurila-et-2012,laurila2013big}. This was a mobile computing research initiative focusing on  generating a deeper scientific understanding of social and behavioral patterns related to mobile technologies. The study took place in Switzerland, and collected various types of longitudinal information including time stamped GPS data from the cell phones of 185 study participants over the course of 18 months. Demographic data such as age and sex is also available. There are approximately 57.5 million GPS location records. The average length of observation for study participants was about 55 weeks. These data are publicly available upon request from the Idiap Research Institute.

\begin{figure}[h]
	\centering
	\includegraphics[height=2.5in]{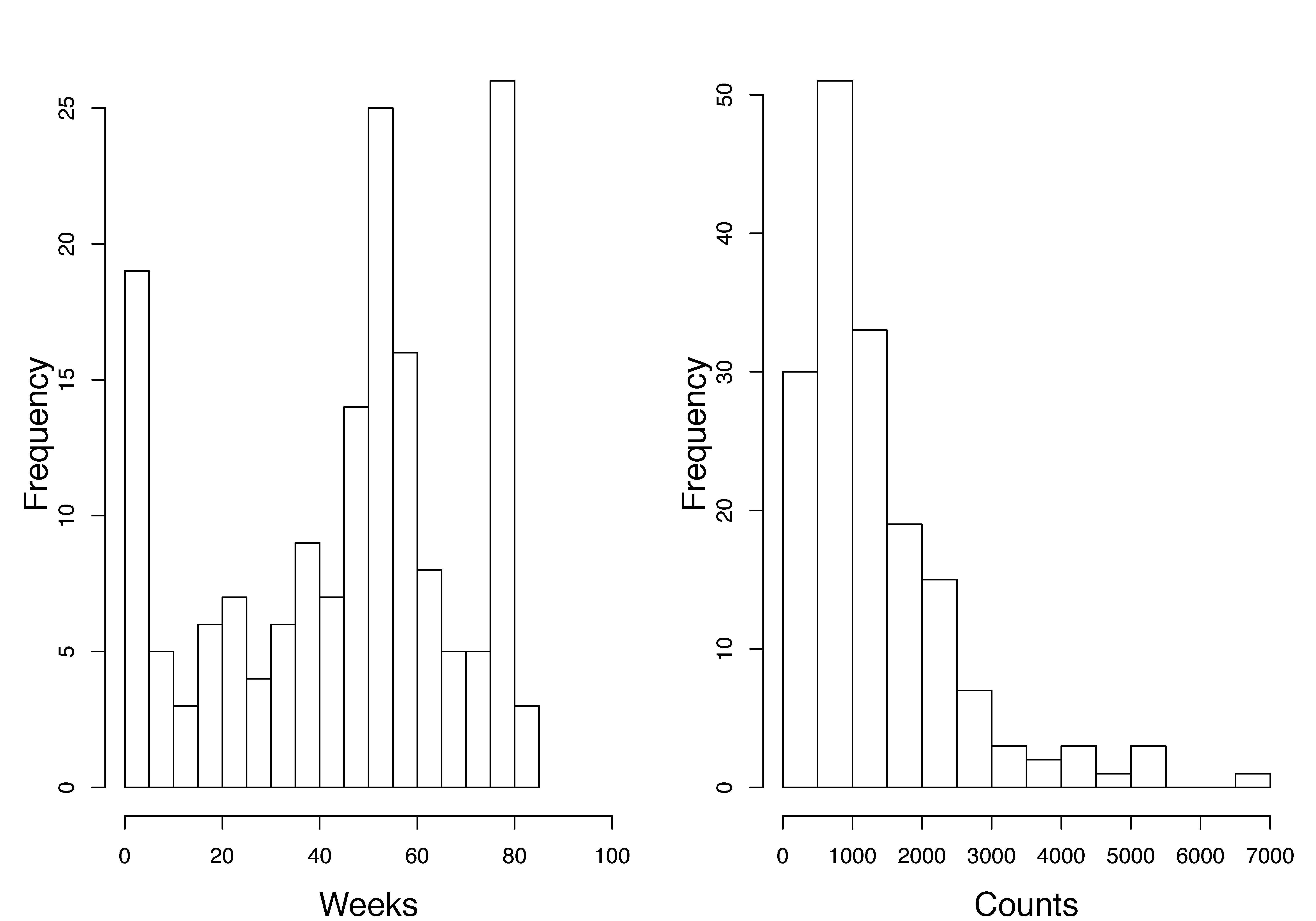}
	\caption{Summary information of the GPS location data. Left panel: histogram of the total length of observation for each study participant expressed in weeks. Right panel: histrogram of the average number of GPS locations per week for each study participant. 
}
	\label{fig::distribution}
\end{figure}

Most activities of daily living of the study participants took place in a rectangular area that we partitioned into $4000^2$ square grid cells with sides of length 28 meters. The locations that do not belong to this spatial observation window were dropped. These locations typically correspond with longer trips took by study participants away from their places of residency. Figure~\ref{fig::distribution} displays summaries of the GPS locations that fall in our chosen spatial observation window. 

For each study participant, we calculated three measures of temporal stability of their mobility patterns: the last crossing time of the average velocity (LCT-velocity) as defined in Eq. \eqref{eq:velocitysample} and Eq. \eqref{eq::LCT}, the last crossing time of the activity distribution (LCT-distribution) as defined in Eq. \eqref{eq::LCT::dis}, and the last crossing time of the level sets of the weekly activity distribution as defined in Eq. \eqref{eq::LCT::LV}. In the calculation of LCT-distribution and LCT-level set, we use the ordinary proportional time estimator defined in Eq. \eqref{eq10}. We used $\alpha = 0.2$ in the determination of level sets, and $\gamma = 0.2$ as the stability threshold for all three measures. The results are summarized in Table \ref{tab:wait}.

\begin{table}
\caption{Means, medians and sample standard deviations of three measures of temporal stability of mobility patterns. The unit of time is weeks.}
	{\begin{tabular}{lccc}\toprule 
		Mobility Measure & Mean & Median & St. Dev. \\
		\midrule
		LCT-velocity & 30.04 & 26 & 17.29\\
		LCT-distribution & 37.18 & 37 & 16.06 \\
		LCT-level set ($\alpha=0.2$) & {17.69} & 17& 9.50\\
		\bottomrule
	\end{tabular}}
	\label{tab:wait}
\end{table}

About 30 weeks of observation is needed until the mobility patterns stabilize according to the LCT-velocity measure. A longer period of time, 37 weeks, is needed until the weekly activity distribution stabilizes. The increased length of the period of observation for this measure is not surprising since it is based on an estimated of the full weekly activity distribution in $N=4000^2$ grid cells. About half of this observation time (18 weeks) is needed to obtain estimates of the $0.2$-level set of the weekly activity distribution which comprise the grid cells in which the study participants spend $80$\% of their weekly time. 

\begin{figure}[h]
	\centering
	\includegraphics[height=2.5in]{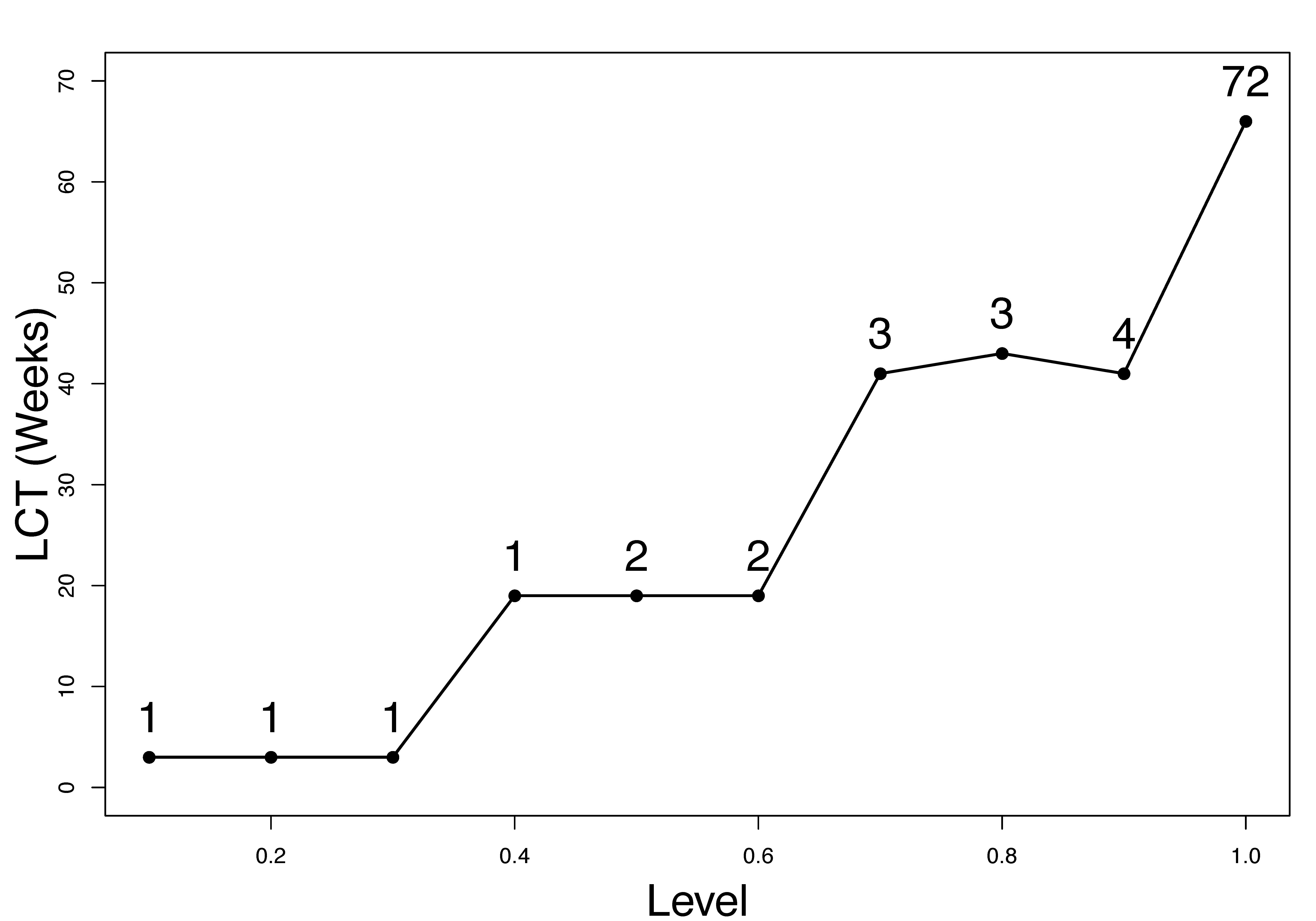}
    \caption{Values of the LCT-level sets $\hat {\sf LCT}_{\sf level, \alpha}(0.2)$ for $\alpha\in \{0.1,0.2,\ldots,1\}$ for an MDC study participant. The unit of time is weeks. The number of connected components of $G_{\sf grid}(L_{\alpha})$ defined by the $\alpha$-level sets $L_{\alpha}$ are shown above the curve. }
	\label{fig::LCT_LV}
\end{figure}

We exemplify how the $\alpha$-level set $L_{\alpha}$ from Eq. \eqref{eq:levelset} and its corresponding LCT-level set $\hat {\sf LCT}_{\sf level, \alpha}(0.2)$ from Eq. \eqref{eq::LCT::LV} change for different values of $\alpha\in [0,1]$. To this end, we define an adjacency graph $G_{\sf grid}$ whose vertices are the $N=4000^2$ grid cells in the spatial observation window. Two grid cells are connected by an edge in $G_{\sf grid}$ if they share an edge or a corner in their arrangement in the spatial observation window \cite{waller-gotway-2004,bivand-et-2013}. We denote by $G_{\sf grid}(L_{\alpha})$ the subgraph of $G_{\sf grid}$ defined by the grid cells in $L_{\alpha}$. We chose a study participant, and determined the level set $L_{\alpha}$, the last crossing time $\hat {\sf LCT}_{\sf level, \alpha}(0.2)$ and the number of connected components of $G_{\sf grid}(L_{\alpha})$ for $\alpha\in \{0.1,0.2,\ldots,1\}$ \--- see Figure~\ref{fig::LCT_LV}. For smaller values of $\alpha$, $L_{\alpha}$ contains grid cells in which the study participant spend the largest proportion of time. When $\alpha \in \{ 0.1,0.2,0.3,0.4\}$, $G_{\sf grid}(L_{\alpha})$ has one connected component which implies that the grid cells that belong to $L_{\alpha}$ are spatially adjacent, and define a single area in which the study participant spends larger amounts of time. The corresponding values of $\hat {\sf LCT}_{\sf level, \alpha}(\gamma)$ are less than 20 weeks which represents the length of observation time needed for reliably detecting this spatial area. For $\alpha \in \{ 0.5, 0.6\}$, $G_{\sf grid}(L_{\alpha})$ has two connected components, and for $\alpha \in\{ 0.7,0.8\}$, $G_{\sf grid}(L_{\alpha})$, $G_{\sf grid}(L_{\alpha})$ has three connected components. Thus this study participant spends their time in grid cells that define two or three spatially contiguous areas. Since these areas include grid cells in which the study participant spends smaller proportions of their weekly time, the length of the observation time needed to identify these areas doubles to about 40 weeks. For $\alpha =1$, $G_{\sf grid}(L_{\alpha})$ has 72 connected components because $L_{\alpha}$ includes grid cells in which the study participant spends very little time. Figure~\ref{fig::LCT_LV} shows that approximately 70 weeks of observation time are needed to detect these grid cells. The same type of plots constructed for other study participants show similar relationships between $\alpha$, $L_{\alpha}$, and $\hat {\sf LCT}_{\sf level, \alpha}(0.2)$.

\begin{figure}
	\centering
	\includegraphics[height=3.5in]{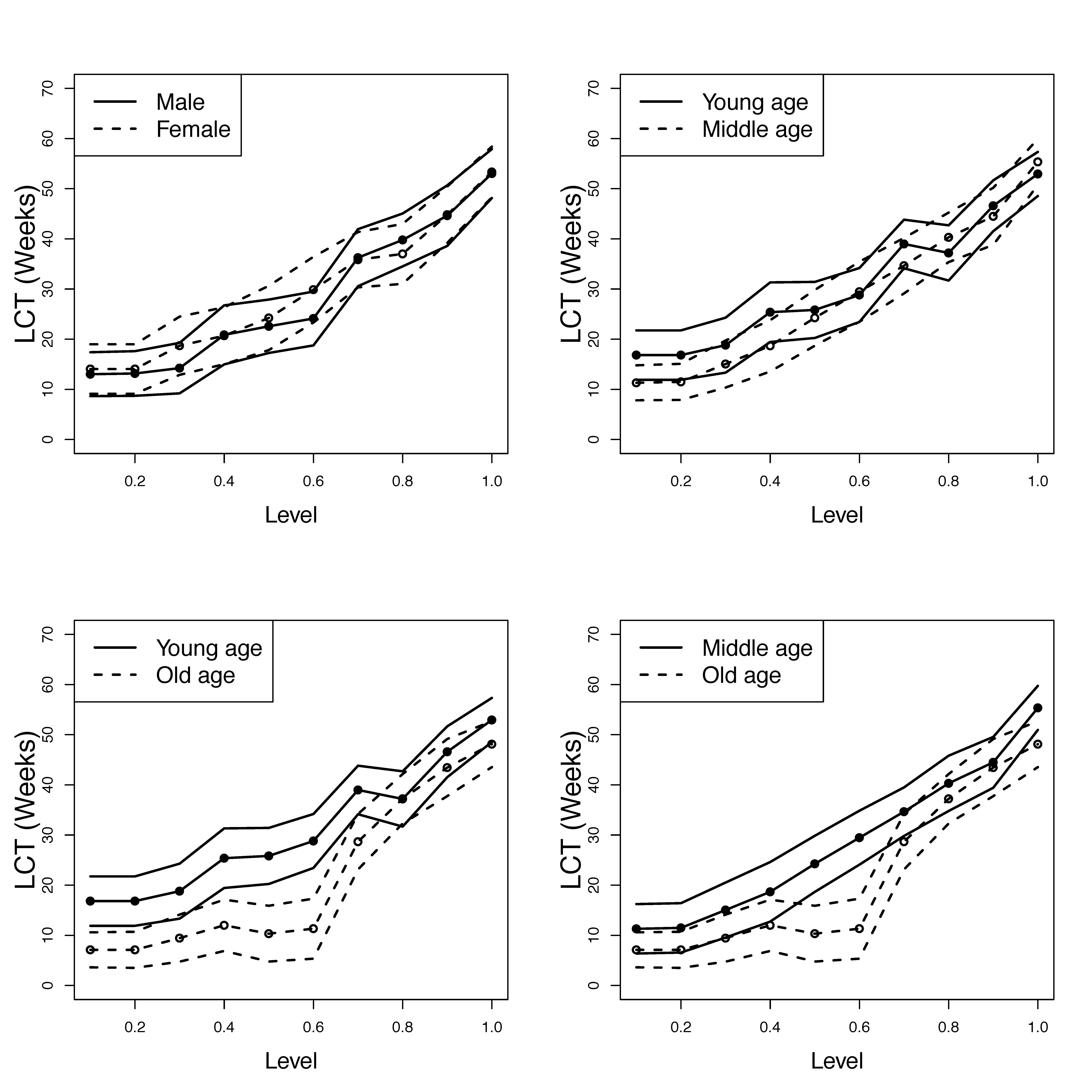}
    	\caption{Mean values and 90\% confidence intervals of the LCT-level sets $\hat {\sf LCT}_{\sf level, \alpha}(0.2)$ for $\alpha\in \{0.1,0.2,\ldots,1\}$ calculated for five demograhic groups: sex (male, female), and age (young, middle, old).}
	\label{fig::LCT_LV::groups}
\end{figure}

Next we want to determine whether the temporal stability of activity distributions varies by the demographic characteristics of the population. We group the study participants by sex (male, female) and age group (young age 15\---34 years old, middle age 35\---54 years old, and old age $\ge$55 years old). For each of these five demographic groups, we calculated the average of the last crossing times of the activity distribution $\hat {\sf LCT}_{\sf level, \alpha}(0.2)$ for every $\alpha\in \{0.1,0.2,\ldots,1\}$. The resulting curves are presented in Figure~\ref{fig::LCT_LV::groups}. The last crossing times at all levels are similar for men and women (see the top left panel). As such, there do not seem to be any sex-based differences in the temporal stability of men and women who live in Switzerland. However, since Switzerland is known to be  a country with very high equality between the two sexes, this finding might not extend to other countries with profound sex inequality. 

In the top right and bottom panels of Figure~\ref{fig::LCT_LV::groups}, we find evidence that the average last crossing times decrease with age especially for levels below $0.5$. This means that mobility patterns are more regular, and consequently are more temporally stable for older study participants compared to younger study participants. The average last crossing times are larger and become very similar across demographic groups for levels above $0.5$ compared to smaller levels below $0.5$. Thus study participants that belong to any of the five demographic groups tend to visit locations they do not typically visit. Longer observation periods are needed to successfully determine these locations. Nevertheless, in order to identify the areas in which study participants spend most of their time, Figure~\ref{fig::LCT_LV::groups} suggests that 10 weeks of observation of GPS locations should suffice for individuals older than $55$. Middle age individuals require about 15 weeks of observation time, while young individuals require about 20 weeks.

\section{Discussion}

The contribution we made in this paper is two fold. On the theoretical side, we proposed the use of last crossing time processes associated with spatiotemporal trajectories of individuals to assess the temporal stability of their mobility patterns. We defined several measures of the temporal dynamics of spatiotemporal trajectories based on the average velocity process, and on human activity distributions in a spatial observation window. We defined the ordinary and the conservative proportional time estimators of human activity distributions, and proved that they are consistent and asymptotically equivalent. We introduced the time period and the ranking time period activity distributions that capture the change in human activity distributions across time periods. We presented related estimators based on GPS location data.

On the empirical side, we analyzed GPS location data collected over a period of 18 months. The previous empirical study \cite{zenk2018many} that focused on assessing the duration of GPS studies is based on data collected over 30 days. By using our new statistical methods and GPS data collected over a much longer period of time, we determined that GPS monitoring needs to be done for at least 15 weeks which represents a minimum study duration about 7 times longer than the 14 days minimum duration recommended in \cite{zenk2018many}. We also put forward the idea that the duration of GPS studies should be assessed by demographic groups. We determined that younger population groups should be monitored for longer periods of time compared to middle age population groups because of their more irregular patterns of mobility. On the other hand, shorter monitoring periods might be needed for older population groups that exhibit mobility patterns that are temporally more stable. We also suggest using our methods to assess the need for different time spans of GPS monitoring for men and women in countries with a known history of inequality between the two sexes. To the best of our knowledge, differential periods of GPS data collection based on demographic groups has not been discussed before. Our work suggests that GPS study designs should take demographic groups into account.

\section*{Funding}

The work of Z.D. and A.D. was partially supported by the National Science Foundation Grant DMS/MPS-1737746 to University of Washington. Y.C. received partial support from the National Science Foundation Grant DMS-1810960 and National Institutes of Health Grant U01-AG016976. The funders had no role in the study design, data collection and analysis, decision to publish, or preparation of the manuscript.

\section*{Acknowledgment}

Portions of the research in this paper used the MDC Database made available by Idiap Research Institute, Switzerland and owned by Nokia.

\appendix

\section{Proofs of theoretical results}

\subsection{Proof of Theorem \ref{thm::equiv}}  \label{proof::equiv}
\begin{proof}

We note that the ordinary proportional time estimator in Eq. \eqref{eq10} can be written as 
\begin{equation}
\hat \pi_{o,j} = \frac{\frac{1}{2}\sum^{n-1}_{i=2}(t_{i+1}-t_{i-1})\mathbf{1}(g_i = G_j)}{\frac{1}{2}(\mathcal{T}+t_n-t_1)},
\label{eq::est::pf::01}
\end{equation}
where $\mathcal{T} = t_{\max}-t_{\min}$. We will first show that the denominators of $\hat \pi_{o,j}$ and $\hat \pi_{c,j}$ are asymptotically the same. Assumption (S2) implies that $\frac{1}{2}(\mathcal{T}+t_n-t_1)\rightarrow \mathcal{T}$, which shows the asymptotic behavior of the denominator of $\hat \pi_{o,j}$. For $\hat \pi_{c,j}$, we have 
\begin{align*}
\sum^n_{i=2}(t_i-t_{i-1})\mathbf{1} (g_i = g_{i-1}) &= \sum^n_{i=2}(t_i-t_{i-1}) - \sum^n_{i=2}(t_i-t_{i-1})\mathbf{1} (g_i \neq g_{i-1}),\\
& = \mathcal{T} -  \sum^n_{i=2}(t_i-t_{i-1})\mathbf{1} (g_i \neq g_{i-1}),\\ 
&\geq \mathcal{T}- M \max_i |t_{i+1}-t_i|, \\
&\rightarrow \mathcal{T},
\end{align*}
where $M$ is the constant from assumption (S3). The limit in the above equation is due to assumption (S1). Thus, the denominators of $\hat \pi_{o,j}$ and $\hat \pi_{c,j}$ are asymptotically the same. Next we focus on the numerators of the two estimators.

The numerator of $\hat \pi_{c,j}$ can be written as 
$$
\sum_{i=2}^n (t_{i+1} - t_i) \mathbf{1}(g_{i+1}=g_i = G_j) = \sum_{i=2}^n A_i,
$$
\noindent where $A_i = (t_{i+1} - t_i) \mathbf{1}(g_{i+1}=g_i = G_j)$. Let $B_i = \frac{t_{i+1}-t_{i-1}}{2} \mathbf{1}(g_{i}=G_j)$. 
Using Eq. \eqref{eq::est::pf::01}, the numerator of $\hat \pi_{o,j}$ can be written as 
$$
\frac{1}{2}\sum^{n-1}_{i=2}(t_{i+1}-t_{i-1})\mathbf{1}(g_i = G_j) = \sum_{i=2}^{n-1}B_i.
$$
When $g_{i-1} = g_{i} = g_{i+1} = G_j$, we have $2B_i = A_i+A_{i-1}$. By assumption (S3), there are at most $2M$ number of time points $t_i$ such that 
the equality $g_{i-1} = g_{i} = g_{i+1} = G_j$ does not hold. Thus 
$$
\sum_{i=2}^{n-1}B_i \mathbf{1}(g_{i-1} = g_{i} = g_{i+1} = G_j)\geq \sum_{i=2}^{n-1}B_i - 2M\cdot \max_i|t_{i+1}-t_i|,
$$
which implies that 
\begin{align}
\hat \pi_{o,j} &\rightarrow \frac{1}{\mathcal{T}}\sum_{i=2}^{n-1}B_i \mathbf{1}(g_{i-1} = g_{i} = g_{i+1} = G_j),\nonumber \\
& = \frac{1}{\mathcal{T}}\sum_{i=2}^{n-1}\frac{A_i + A_{i-1}}{2}\mathbf{1}(g_{i-1} = g_{i} = g_{i+1} = G_j). \label{eq:pi0limit}
\end{align}
Again, using the fact that there are at most $2M$ number of time points $t_i$ such that 
the equality $g_{i-1} = g_{i} = g_{i+1} = G_j$ does not hold, we obtain
\begin{align*}
\sum_{i=2}^{n-1}A_i\mathbf{1}(g_{i-1} = g_{i} = g_{i+1} = G_j) &\geq \sum_{i=2}^{n}A_i -(2M+1) \cdot \max_i|t_{i+1}-t_i|,\\
\sum_{i=2}^{n-1}A_{i-1}\mathbf{1}(g_{i-1} = g_{i} = g_{i+1} = G_j) &\geq \sum_{i=2}^{n}A_i -(2M+1) \cdot \max_i|t_{i+1}-t_i|.
\end{align*}
It follows that 
\begin{align*}
\hat \pi_{c,j}& = \frac{\sum^n_{i=2}(t_i-t_{i-1})\mathbf{1} (g_i = g_{i-1} = G_j)}{\sum^n_{i=2}(t_i-t_{i-1}) \mathbf{1} (g_i = g_{i-1})}\\
&\rightarrow \frac{1}{\mathcal{T}}\sum_{i=2}^n A_i,\\
&\rightarrow \frac{1 }{\mathcal{T}}\sum_{i=2}^{n-1}\frac{A_i+A_{i-1}}{2}\mathbf{1}(g_{i-1} = g_{i} = g_{i+1} = G_j),
\end{align*}
which is the same limit in Eq. \eqref{eq:pi0limit} we obtained for $\hat \pi_{o,j}$. Therefore the numerators of $\hat \pi_{o,j}$ and $\hat \pi_{c,j}$ are asymptotically the same, which proves that $\hat \pi_{o,j}$ and $\hat \pi_{c,j}$ are asymptotically equal.

\end{proof}

\subsection{Proof of Theorem \ref{thm::convergencerule}} \label{proof:convergencerule}
\begin{proof}

Theorem~\ref{thm::equiv} proves that the two estimators are asymptotically equivalent. Thus, we only need to derive the convergence of one of the two estimators  to the true activity distribution $\pi=(\pi_1,\ldots,\pi_N)$ from Eq. \eqref{eq:activityunif}. In what follows we focus on the conservative proportional time estimator.

Without loss of generality, we assume that there exist $K\ge 1$ disjoint time intervals in which the individual is inside grid cell $G_j$, i.e.,
there are $[a_1,b_1],\cdots, [a_K,b_K]$ such that $a_i< b_{i}<a_{i+1}$ for $i=1,\ldots,K-1$, $t_{min}\le a_1$, $b_K\le t_{\max}$ and 
$$
\{t: G(t)\in G_j\} = [a_1,b_1]\cup\cdots \cup [a_K, b_K].
$$
Since, in the definition of the true activity distribution $\pi$, $T$ follows a uniform distribution on the reference time frame $[t_{\min},t_{\max}]$, we can express $\pi_j$ as 
$$
\pi_j = \P (G(T)\in G_j) = \sum_{k=1}^K \P (T\in [a_k,b_k]) = \frac{1}{\mathcal{T}}\sum_{k=1}^K (b_k-a_k).
$$
As before, $\mathcal{T} = t_{\max}-t_{\min}$.

For the interval $[a_k,b_k]$, we let $t_{i_*}$ be the first observation time after $a_k$, and $t_{i_{**}}$ be the last observation time before $b_k$:
$$
t_{i_*}\geq a_k,\quad t_{i_*-1}<a_k,\quad t_{i_{**}+1}>b_k,\quad t_{i_{**}}\leq b_k.
$$
Because $G(t)\in G_j$ for all $t\in [a_k,b_k]$, we have $g_i \in G_j$ for all $i\in \{i_*,i_*+1,\ldots,i_{**}\}$. 
The conservative proportional time estimator estimates the length of the interval $[a_k,b_k]$ based on the length of the interval $[t_{i_*},t_{i_{**}}]$. The corresponding error is
\begin{align*}
|(b_k-a_k) - (t_{i_{**}} - t_{i_*})| &\leq t_{i_*}-a_k + b_k-t_{i_{**}},\\
& \leq (t_{i_*} - t_{i_*-1}) + (t_{i_{**}+1} - t_{i_{**}}),\\
&\leq 2\max\limits_{i=1,\ldots,n-1} |t_{i+1}-t_i|\rightarrow 0,
\end{align*}
due to assumption (S1).

By applying the above argument to each interval $[a_k,b_k]$, $k=1,\ldots,K$, we conclude that 
$$
\sum_{i=2}^n (t_{i}-t_{i-1})\mathbf{1}(g_i = G_j) \rightarrow \sum_{k=1}^K (b_k-a_k) . 
$$
Because 
$$
\sum_{i=2}^n (t_{i}-t_{i-1})\mathbf{1}(g_i= G_j)\geq \sum_{i=2}^n (t_{i}-t_{i-1})\mathbf{1}(g_i=g_{i-1} = G_j) - M\cdot \max\limits_{i=1,\ldots,n-1} |t_{i+1}-t_i|,
$$
we further conclude that 
$$
\sum_{i=2}^n (t_{i}-t_{i-1})\mathbf{1}(g_i=g_{i-1} = G_j)   \rightarrow \sum_{k=1}^K (b_k-a_k).
$$
This proves the convergence of the conservative proportional estimator to the true activity distribution:
\begin{align*}
\hat \pi_{c,j} & \rightarrow \frac{\sum_{i=2}^n (t_{i}-t_{i-1})\mathbf{1}(g_i=g_{i-1} = G_j)}{\mathcal{T}},\\
&\rightarrow \frac{\sum_{k=1}^K (b_k-a_k)}{\mathcal{T}},\\
& = \pi_j. 
\end{align*}

\end{proof}

\appendix

%\bibliographystyle{tfs}
%\bibliography{references}

\end{document}